\documentclass[12pt]{article}

\usepackage{cite}
\usepackage{graphics,epsfig}

\usepackage{amssymb,latexsym}

\usepackage{amsmath,amssymb,amsfonts}	
\usepackage{color}
\usepackage{amsmath}
\usepackage{amsfonts}
\usepackage{amssymb}
\usepackage{graphicx}
\usepackage{slashed}            
\usepackage{bbm}                
\usepackage{xspace}				

\usepackage{mathrsfs}

\usepackage{tikz}
\usetikzlibrary{arrows,shapes}
\usetikzlibrary{trees}
\usetikzlibrary{matrix,arrows} 				
\usetikzlibrary{positioning}				
\usetikzlibrary{calc,through}				
\usetikzlibrary{decorations.pathreplacing}  
\usepackage{pgffor}							

\usetikzlibrary{decorations.pathmorphing}	
\usetikzlibrary{decorations.markings}
\tikzset{
    vector/.style={decorate, decoration={snake}, draw},
	provector/.style={decorate, decoration={snake,amplitude=2.5pt}, draw},
	antivector/.style={decorate, decoration={snake,amplitude=-2.5pt}, draw},
    fermion/.style={draw=black, postaction={decorate},
        decoration={markings,mark=at position .55 with {\arrow[draw=black]{>}}}},
    fermionbar/.style={draw=black, postaction={decorate},
        decoration={markings,mark=at position .55 with {\arrow[draw=black]{<}}}},
    fermionnoarrow/.style={draw=black},
    gluon/.style={decorate, draw=black,
        decoration={coil,amplitude=4pt, segment length=5pt}},
    scalar/.style={dashed,draw=black, postaction={decorate},
        decoration={markings,mark=at position .55 with {\arrow[draw=black]{>}}}},
    scalarbar/.style={dashed,draw=black, postaction={decorate},
        decoration={markings,mark=at position .55 with {\arrow[draw=black]{<}}}},
    scalarnoarrow/.style={dashed,draw=black},
    electron/.style={draw=black, postaction={decorate},
        decoration={markings,mark=at position .55 with {\arrow[draw=black]{>}}}},
	bigvector/.style={decorate, decoration={snake,amplitude=4pt}, draw},
}

\tikzstyle{block} = [draw, rectangle, 
    minimum height=3em, minimum width=6em]

\usepackage{hyperref}			

\DeclareMathAlphabet{\mathpzc}{OT1}{pzc}{m}{it}

\let\a=\alpha \let\b=\beta \let\g=\gamma \let\d=\delta \let\e=\epsilon
  \let\th=\theta  \let\k=\kappa
\let\l=\lambda \let\m=\mu \let\n=\nu \let\x=\xi \let\p=\pi 
\let\s=\sigma   \let\f=\phi  
 
      \let\G=\Gamma  \let\Th=\Theta 
\let\X=\Xi  \let\S=\Sigma  \let\Y=\Psi
 
\let\la=\label  
  
\def\nn{\nonumber} \def\bd{\begin{document}} \def\ed{\end{document}}
\def\ds{\documentstyle} \let\fr=\frac \let\bl=\bigl \let\br=\bigr
\let\Br=\Bigr \let\Bl=\Bigl
\let\bm=\bibitem
\let\na=\nabla
\def\tU{{\widetilde U}}
\let\pa=\partial \let\ov=\overline
\def\ie{{\it i.e.\ }}
\newcommand{\be}{\begin{equation}}
\newcommand{\ee}{\end{equation}}
\def\ba{\begin{array}}
\def\ea{\end{array}}
\def\ft#1#2{{\textstyle{{\scriptstyle #1}\over {\scriptstyle #2}}}}
\def\fft#1#2{{#1 \over #2}}
\def\F#1#2{{ F_{#1}^{(#2)} }}
\def\cF#1#2{{ {\cal F}_{#1}^{(#2)} }}

\def\R{{\bf R}}
\def\sst#1{{\scriptscriptstyle #1}}
\def\oneone{\rlap 1\mkern4mu{\rm l}}
\def\e7{E_{7(+7)}}
\def\td{\tilde}
\def\wtd{\widetilde}
\def\im{{\rm i}}
\def\bog{Bogomol'nyi\ }
\newcommand{\ho}[1]{$\, ^{#1}$}
\newcommand{\hoch}[1]{$\, ^{#1}$}
\newcommand{\bea}{\begin{eqnarray}}
\newcommand{\eea}{\end{eqnarray}}
\newcommand{\ra}{\rightarrow}
\newcommand{\lra}{\longrightarrow}
\newcommand{\Lra}{\Leftrightarrow}
\newcommand{\ap}{\alpha^\prime}
\newcommand{\bp}{\tilde \beta^\prime}
\newcommand{\cB}{{\cal B}}
\newcommand{\cO}{{\cal O}}
\newcommand{\vecx}{\vec{x}}
\newcommand{\vecy}{\vec{y}}
\newcommand{\vecp}{\vec{p}}
\newcommand{\vecq}{\vec{q}}
\newcommand{\tr}{{\rm tr} }
\newcommand{\Tr}{{\rm Tr} }
\newcommand{\NP}{Nucl. Phys. }

\newcommand{\cL}{{\cal L}}
\newcommand{\cA}{{\cal A}}
\newcommand{\cT}{{\cal T}}
\newcommand{\cD}{{\cal D}}
\newcommand{\cH}{{\cal H}}

\def\sst#1{{\scriptscriptstyle #1}}
\def\0{{\sst{(0)}}}
\def\1{{\sst{(1)}}}
\def\2{{\sst{(2)}}}
\def\3{{\sst{(3)}}}
\def\4{{\sst{(4)}}}
\def\5{{\sst{(5)}}}
\def\6{{\sst{(6)}}}
\def\7{{\sst{(7)}}}
\def\8{{\sst{(8)}}}
\def\9{{\sst{(9)}}}
\def\p{{\sst{(p)}}}
\def\q{{\sst{(q)}}}
\def\ve{\varepsilon}
\def\vf{\varphi}
\def\F{\Phi}
\def\wg{\wedge}

\def\thb{\bar{\theta}}
\def\Thb{\bar{\Theta}}
\def\barp{\bar{p}}
\def\barq{\bar{q}}
\def\barc{\bar{c}}
\def\bard{\bar{d}}
\def\e{\epsilon}

\def \bi{\bibitem}
\def \la {\label}

\def \l {\lambda}
\def\foot{\footnote}
\def \tl  {{\tilde \l}}
\def \sql {{\sqrt \l}}
\def \adss {$AdS_5 \times S^5$\ }
\newcommand{\rf}[1]{(\ref{#1})}
\def \ov {\over}

\def\th{\theta}
\def\Th{\Theta}
\def\vth{\vartheta}
\def\btheta{{\bar\theta}}
\def\ttheta{{{\tilde\theta}}}
\def\bttheta{{{\bar\ttheta}}}
\def\vth{\vartheta}

\def\ra{\rightarrow}
\def\N{\nabla}
\def\F{{\cal F}}
\def\uM{\underline{M}}
\def\uA{\underline{A}}
\def\uN{\underline{N}}
\def\uP{\underline{P}}
\def\ua{\underline{a}}
\def\ub{\underline{b}}
\def\uc{\underline{c}}
\def\ud{\underline{d}}
\def\ue{\underline{e}}
\def\uf{\underline{f}}
\def\ui{\underline{i}}
\def\uj{\underline{j}}
\def\uk{\underline{k}}
\def\ul{\underline{l}}
\def\ual{\underline{\alpha}}
\def\ube{\underline{\beta}}
\def\um{\underline{m}}
\def\un{\underline{n}}
\def\up{\underline{p}}
\def\uq{\underline{q}}
\def\ur{\underline{r}}
\def\us{\underline{s}}
\def\umu{\underline{\mu}}
\def\unu{\underline{\nu}}
\def\ula{\underline{\l}}
\def\uka{\underline{\k}}
\def\usi{\underline{\s}}
\def\urh{\underline{\r}}
\def\cc{\circ}
\def\eqv{\equiv}

\def\ni{\noindent}

\def\Ep{E^{{}^{(+)}}}
\def\Em{E^{{}^{(-)}}}

\def\Mp{M^{{}^{(+)}}}
\def\Mm{M^{{}^{(-)}}}

\def \ha{{1\ov 2}}

\def\r{\rho}

\def\Y{{\rm Y}}
\def\X{{\rm X}}
\def\tY{\tilde{\rm Y}}
\def\tX{\tilde{\rm X}}
\def\dY{\dot{\rm Y}}
\def\dX{\dot{\rm X}}

\def \J {\mathcal{J}}
\def \del {\partial}

\def\dF{\dot{F}}
\def\dG{\dot{G}}
\def\df{\dot{f}}
\def \E {{\cal E}}
\def \S {{\cal S}}
\def \J {{\cal J}}

\def\ms{\mathcal{S}}
\def\mj{\mathcal{J}}
\def\soj{\fr{\ms}{\mj}}
\def \R {{\bf R}}
\def \om {\omega}
\def \bE {\bar E}
\def \x {{\cal X}}

\def \bi{\bibitem}
\def \la {\label}

\def \l {\lambda}
\def\foot{\footnote}
\def \tl  {{\tilde \l}}
\def \sql {{\sqrt \l}}
\def \adss {$AdS_5 \times S^5$\ }
\def \ov {\over}

\def \varpi {{\rm w}}

\def\thb{\bar{\theta}}
\def\Thb{\bar{\Theta}}
\def\mb{\bar{\m}}
\def\ab{\bar{\a}}
\def\zb{\bar{z}}
\def\psib{\bar{\psi}}
\def\barp{\bar{p}}
\def\barq{\bar{q}}
\def\barc{\bar{c}}
\def\bard{\bar{d}}
\def\e{\epsilon}
\def\wb{\bar{w}}
\def\lb{\bar{\l}}
\def\Jb{\bar{J}}
\def\Nb{\bar{N}}
\def\Zb{\bar{Z}}
\def\pab{\bar{\pa}}

\def\Cb{\bar{C}}

\def\At{\tilde{A}}
\def\Bt{\tilde{B}}
\def\Ct{\tilde{C}}
\def\Dt{\tilde{D}}
\def\Et{\tilde{E}}
\def\Ft{\tilde{F}}
\def\Gt{\tilde{G}}
\def\Ht{\tilde{H}}
\def\Kt{\tilde{K}}
\def\Mt{\tilde{M}}
\def\Nt{\tilde{N}}
\def\Rt{\tilde{R}}
\def\at{\tilde{a}}
\def\bt{\tilde{b}}
\def\ct{\tilde{c}}
\def\dt{\tilde{d}}
\def\et{\tilde{e}}
\def\ft{\tilde{f}}
\def\htil{\tilde{h}}
\def\gt{\tilde{g}}
\def\nt{\tilde{n}}
\def\mut{\tilde{\mu}}
\def\nut{\tilde{\nu}}
\def\pht{\tilde{\f}}
\def\vft{\tilde{\vf}}

\def\rht{\tilde{\rho}}

\def\asth{\hat{*}}
\def\phh{\hat{\phi}}

\def\bA{{\bf A}}

\def\ola{\overleftarrow}
\def\ora{\overrightarrow}
\def\alt{\tilde{\a}}

\def\eh{\hat{e}}
\def\eph{\hat{\e}}
\def\ph{\hat{p}}
\def\alh{\hat{\a}}
\def\beh{\hat{\b}}
\def\gah{\hat{\g}}
\def\Fh{\hat{F}}
\def\muh{\hat{\m}}
\def\nuh{\hat{\n}}
\def\thh{\hat{\th}}
\def\rhh{\hat{\r}}
\def\dh{\hat{d}}
\def\ih{\hat{i}}
\def\jh{\hat{j}}
\def\hh{\hat{h}}
\def\nh{\hat{n}}
\def\gh{\hat{g}}
\def\kh{\hat{k}}
\def\deh{\hat{\d}}
\def\wh{\hat{w}}
\def\lah{\hat{\l}}
\def\Ah{\hat{A}}
\def\Kh{\hat{K}}
\def\Nh{\hat{N}}
\def\Rh{\hat{R}}
\def\Ch{\hat{C}}
\def\Omh{\hat{\Omega}}

\def\xh{\hat{x}}

\def\ps{\rlap{\, /}\;\,p }
\def\ks{\rlap{\, /}\;\,k }

\def\gym{g_{YM}}

\def\adot{\dot{a}}
\def\bdot{\dot{b}}
\def\bpa{\bar{\pa}}

\def\pr{\prime}
\def\ssk{\medskip}
\def\clb{\color{blue}}
\def\clr{\color{red}}
\def\clg{\color{green}}

\begin{document}

\overfullrule=0pt
\parskip=2pt
\parindent=12pt
\headheight=0in \headsep=0in \topmargin=0in
\oddsidemargin=0in

\vspace{ -3cm}
\thispagestyle{empty}

 \vspace{0.1cm}

\setcounter{equation}{0}
\setcounter{footnote}{0}
\setcounter{section}{0}


\begin{center}
{\Large\bf Hypersurface foliation approach to renormalization of ADM formulation of gravity}

\medskip
\bigskip\color{black}\vspace{0.6cm}
{
{I. Y. Park}
}
\\[7mm]
{\it Department of Physics, Hanyang University \\
Seoul 133-791, Korea}\\

\vspace{0.3cm}
{\it Department of Applied Mathematics,
Philander Smith College 
                               \\
Little Rock, AR 72223, USA \\
inyongpark05@gmail.com
}
\bigskip

\bigskip\bigskip

{
\centerline{\large\bf Abstract}
\begin{quote}

We carry out ADM splitting in the Lagrangian formulation and establish a procedure in which (almost) all of the
unphysical components of the metric are removed by using the 4D diffeomorphism and the measure-zero 3D symmetry. The procedure introduces a constraint that corresponds to the Hamiltonian constraint of the Hamiltonian formulation, and its solution implies that the 4D dynamics admits an effective description through 3D hypersurface physics. As far as we can see, our procedure implies potential renormalizability of {the ADM formulation of} 4D Einstein gravity for which {a complete gauge-fixing} in the ADM formulation and hypersurface foliation of geometry are the key elements. 
If true, this implies that the alleged unrenormalizability of 4D Einstein gravity may be due to the presence of the unphysical fields.
{The procedure can straightforwardly be applied to quantization around a flat background; the Schwarzschild case seems more subtle.} We discuss a potential limitation of the procedure when applying it to explicit time-dependent backgrounds.

\end{quote}}

\end{center}


\newpage

\section{Introduction}

The quantization of 4D Einstein-Hilbert action has been a long-standing problem. (See \cite{Carlip:2001wq} for a review.)
One-loop renormalizability of pure 4D gravity (i.e., gravity without any matter field coupled) was established in \cite{'tHooft:1974bx}. However, the presence of matter fields does not preserve the one-loop renormalizability \cite{Deser:1974cy}\cite{Deser:1974cz}. Also, it was subsequently shown \cite{Goroff:1985th} that two-loop and higher order diagrams of the pure gravity require proliferation of counter terms, thereby leading to unrenormalizability.
Needless to say, the lack of renormalizability has been a serious obstacle that has delayed (or even blocked) progress in many fundamental issues such as the black hole information paradox.

There have been several approaches to quantization of gravity. The first was through the Hamiltonian formulation of gravity \cite{Arnowitt:1962hi}\cite{Gourgoulhon:2007ue}. (Earlier discussions can be found, e.g., in \cite{Dirac:1958jc,Bergmann:1972ud,Isham:1984sb}.)  The obstacle in this approach was the complexity of Hamiltonian constraint; a different approach based a different set of variables was proposed and is now known as loop quantum gravity \cite{Ashtekar:1986yd,Rovelli:1997yv,Thiemann:2007zz}. Still another approach is based on adding higher derivative terms in the action \cite{Stelle:1976gc,Antoniadis:1986tu}.

In this work, we show by conducting the 3+1 splitting in the Lagrangian formulation that
the dynamics of the pure 4D gravity is effectively reduced\footnote{To our pleasant surprise, we recently have found the work of \cite{Fischer:1996qg} where a Hamilatonain reduction led to lower dimensional volume. There also is an earlier related work \cite{York:1972sj}. We will comment on these further in the conclusion.} to the dynamics of pure ``3D gravity." (As well known, the genuine 3D gravity does not have a graviton.
Our ``3D" theory - originating from 4D gravity - is not the genuine 3D gravity; as we will see, it has two propagating degrees of freedom that are inherited from 4D. It holographically represents the original 4D system.) Since pure 3D gravity is known to be renormalizable \cite{Witten:1988hc}\cite{Anselmi:2003xb}, our result implies that the 4D Einstein gravity is renormalizable: renormalizability of the 4D gravity is achieved essentially by removing all of the unphysical degrees of
freedom in the particular manner which we will describe in detail in the main body. Some of the analyses in this work have been repeated from different angles in \cite{Park:2014noa,Park:2015qxa,Park:2015ota}.

{It may be good to clearly state at this point the task undertaken in the present work. The equivalence between the usual formulation and ADM formulation of general relativity was questioned in \cite{Kiriushcheva:2008sf}. In our view, it is a legitimate concern given, e.g., that the ADM formulation may not generally be applicable to an arbitrary spacetime but is most useful in dealing with a globally hyperbolic spacetime. The ADM formulation should be applicable to a ``locally hyperbolic" spacetime when the issue under consideration is local in nature such as renormalizability. Furthermore, once one takes the gauge in which the shift vector is set to zero, the analysis becomes essentially that of the usual formulation in the synchronous gauge.		 It is certainly possible to repeat all of the analysis in this work without referring to the ADM formulation but rather within the usual Lagrangian and Hamiltonian formulations with the synchronous gauge.
Since gauge-fixing is a subtle issue in general, we clearly state the task undertaken in this work: quantization associated with the physical states dictated by the ADM formulation with the synchronous type gauge.}

 The focus of the present work is the possibility that the alleged unrenormalizability may be due to the presence of unphysical degrees of freedom in the 4D Einstein action. There are eight unphysical degrees of freedom, and four of them are associated with the 4D diffeomorphism.
The other four degrees of freedom are non-dynamical, and may explicitly be removed. In principle, there is a chance that such removal may eliminate the need for some of the counter terms, and we confirm this anticipation in a dramatic way.

{Basically, we remove the unphysical fields by gauge fixings. (It may appear that the procedure depends on the gauge-fixing too sensitively. We will have more on this in the conclusion.) The field equations of the lapse function and shift vector do not have any time derivative acting on them. {\em Therefore, it should be unnecessary to use the bulk gauge symmetry to gauge-fix them. Instead, one may use the residual symmetry after the bulk fixing to gauge-fix these non-dynamic fields.} (This idea may be related to the one in \cite{Isham:1984sb}.)
Gauge-fixing of the bulk diffeomorphism will be carried out in the standard manner; it is the gauge-fixing of the non-dynamical fields that makes the difference.\footnote{In this work we adopt the approach in which all of the unphysical degrees of freedom are explicitly fixed. Alternatively one may keep the off-shell unphysical degrees of freedom that are associated with the lapse and shift. See \cite{Park:2015ota} for this approach.} In terms of Dirac's terminology \cite{Dirac}, the non-dynamical fields introduce the first class constraints in the Hamiltonian formalism. 
 In the spirit of the comment made in section 7.6 of \cite{Weinberg1} (in which it was stated that the first class constraints can be eliminated by a choice of gauge), we deal with the Lagrangian analogue of the first class constraints simply by additional gauge-fixings.
 In the Hamiltonian approach, the first class constraints are associated with gauge symmetries. The necessity of such a gauge-fixing was discussed in \cite{Woodard:1989ac}\cite{Carlip:2001wq}. Since the non-dynamical fields do not evolve in time, and are thus virtually three-dimensional, they should be viewed as generating three-dimensional (as opposed to four-dimensional) symmetries. Because of this, a measure-zero symmetry will be used to remove the non-dynamical fields.}
 { (See \cite{Park:2014noa} for an explicit and quantitative analysis on the nature of the residual symmetry { and ghost-term related issues}.)}

{In spite of the unphysical nature of the non-dynamical fields, the conventional procedure includes them in the renormalization program: they appear as external lines of loop diagrams in the Green function computations. They also run around the loops of various loop diagrams. The main reason 
for not gauge-fixing them should be maintaining the covariance. (Moreover, such a procedure was successful in the cases of gauge theories.) However, if there exists a (non-covariant) renormalization procedure that does not include the non-dynamical fields and is still controllable, it would be worthwhile to examine the procedure. We show that there indeed exists such a procedure that maintains 3D covariance.}

Since we claim in this work that a different (and complete) gauge-fixing may render the 4D gravity renormalizable,\footnote{The analogous phenomenon does not occur in gauge theories since gauge theories are renormalizable in the Lorentz
gauge that keeps the non-dynamical time component. It seems that gravity is just different from gauge theories in this aspect.} one may raise a question how that could happen given that physics should not depend on a gauge choice. Firstly, let us recall that only the correlator of gauge invariant quantities are independent of the gauge-fixing; in a renormalization procedure, one considers correlators of the elementary fields - which may or may not be gauge singlets (for example, the metric is not a gauge singlet). Therefore, renormalization procedure {\em does} depend on the gauge choice, and we have demonstrated this in the appendix. What we observe in this work is that the dependence can be drastic. A second clue to the answer to the question above can be found in \cite{Mazur:1989by}. We will come back to this work later but briefly the authors noted that the presence of an unphysical mode makes the path-integral divergent and  ill-defined. The divergence here is something that one should worry about before one examines renormalizability of the theory, and is worse in that sense than unrenormalizability. Therefore, it seems reasonable to expect that removing all, as opposed to four (as in, e.g., de Donder gauge) out of eight, of the unphysical modes can, in principle, make a drastic difference. Therefore, gauge-choice independence would be expected for different gauges as far as they are complete fixings.

There exists a gauge, called the radiation gauge \cite{Smarr:1978dia}\cite{Wald}, in which all of the unphysical fields are removed. The gauge-fixing procedure that we adopt keeps only the physical degrees of freedom as in the radiation gauge, but is different from the radiation gauge in that it leads to effective reduction of the dynamics onto the hypersurface. The hypersurface foliation approach\footnote{The hypersurface foliation approach combined with explicit dimensional reduction has been fruitful \cite{Sato:2002kv,Park:2013iqa,Park:2013bma,Park:2014mba}. Unlike these works, no explicit dimensional reduction is carried out here: the projection onto 3D is dictated by removal of unphysical degrees of freedom through the measure-zero diffeomorphism. Thus the reduction is ``spontaneous".} in this work provides a convenient arena for examining the renormalizability after removal of the unphysical degrees of freedom.
Let us consider the 3+1 splitting of the 4D Einstein-Hilbert action.
Compared with the existing covariant approach, the ADM formulation employed in this manuscript has two advantages:
Firstly, the ADM formulation is effective in organizing the degrees of freedom for easy isolation of the unphysical degrees of freedom. In other words, the formulation readily identifies the non-dynamical fields thereby setting the stage for their removal.
Secondly, the ADM splitting brings out the utility of the measure-zero gauge symmetry for removal of the non-dynamic fields. This feature will play a crucial role in section 2.

\vspace{.3in}
{The rest of the paper is organized as follows. 
In section 2, we discuss removal of unphysical degrees of freedom. 
After starting off by recalling the gauge-fixing procedure of a YM type gauge theory, we note that it should be possible to gauge-fix the lapse function and shift vector by the measure-zero diffeomorphisms after fixing the bulk gauge symmetry (de Donder gauge will be adopted for the bulk fixing). As well known (see, e.g., \cite{Poisson}), the choice of the lapse and shift is arbitrary, hence gauge-fixing the lapse and shift should be a legitimate procedure in any case. What is important is that it should be possible to gauge-fix them by using the measure-zero gauge symmetry, not the bulk ones. The result of these gauge-fixings is the projection of 4D dynamics onto the 3D hypersurface: the 3D system has two physical degrees of freedom inherited from 4D. (The reduction is limited to a pure Einstein system; a matter field, if present, will not be reduced, at least not in any simple way.) By invoking the logic of \cite{Goroff:1985th} and \cite{Anselmi:2003xb} one arrives at renormalizability. This will be pointed out in section 3, in which we will also comment on precisely what physics the reduced theory should describe.
The procedure should be viewed as a generalization of the holography idea of \cite{'tHooft:1993gx}.\footnote{'t Hooft observed the Holography in the black hole context. If what we propose here is true, gravity theory itself has the holographic property through a lower dimensional gravity theory.} A flat spacetime will be considered throughout. { It should be possible to apply the procedure to a Schwarzschild black hole background with relatively minor modifications.} However, it is not clear whether the procedure can be applied to a more complex background such as an explicit time-dependent black hole background, and this is  potentially a limitation of the procedure.
For the case of globally hyperbolic spacetimes, the reduction can easily be understood from a different and more mathematical perspective \cite{Park:2014qoa}. (The condition of global hyperbolicity is not strictly required though.) This result is summarized in section 3.2. In Conclusion, we comment on several issues such as recovery of the 4D covariance in the present context or the fully 4D covariant formulation of the quantization. (Progress has been recently made in \cite{Park:2015ota}.) 
 In Appendix, we illustrate the gauge-fixing dependence of a renormalization procedure by taking a system of metric coupled with a scalar. The sole purpose of considering this system is to demonstrate the dependence: we consider only the pure gravity system in the main body. }

\section{Removal of unphysical degrees of freedom}

It will be useful for what follows to recall the quantization procedure in Maxwell's theory. The vector field has four
components to start with but only two of them are physical degrees of freedom. The system has gauge symmetry; it reduces the number of degrees of freedom to three. The time component is non-dynamical, leading to the further reduction of the number of physical fields to two. 
Let us consider temporal gauge to be specific. It turns out that temporal gauge does not entirely use up the gauge freedom but leaves gauge symmetry associated with the hypersurface of a fixed time \cite{Girotti:1984gq}.
For the perspective of our gravity analysis, what is important is that the non-dynamical time component can be gauge-fixed without using the full bulk symmetry but instead by using measure-zero lower dimensional gauge symmetry. Below we will show that there is an analogous procedure in general relativity.

As in the Maxwell's case, it is the close conspiracy between non-dynamism and { gauge symmetry} that brings complete removal of the unphysical degrees of freedom.
In due course of the analysis below, the lapse function $n$ will get fixed to $n=1$. (This choice is suitable because we are considering expansion of the theory around a flat spacetime to be specific.) This introduces a constraint that corresponds to Hamiltonian constraint of the Hamiltonian formulation. As we will see, the constraint can be solved, and its solution implies, among other things, the effective projection of the 4D dynamics onto the 3D hypersurface.

{The gauge-fixings of the measure-zero symmetries will be carried out following the spirit of section 15.4 of \cite{Weinberg2} in which the first class constraint was eliminated by fixing the corresponding symmetry. In essence, what we do here is fix the gauge and explicitly solve the resulting constraints. Therefore the procedure does not introduce any ghosts at the bulk level. (In \cite{Weinberg2}, quantization in the axial gauge was analyzed, and it was shown that the axial gauge quantization does not introduce any ghosts. For the actual perturbation computations, a covariant gauge -  
 which does introduce ghosts - was used. { See \cite{Park:2014noa} for the ghost-related issues.}) }

\subsection{Isolation of unphysical degrees of freedom }

Consider the 4D Einstein-Hilbert action (see, e.g., \cite{Straumann} for a review)
\bea
S=\int d^4 x \sqrt{-\gh}\;R  \la{unsplit}
\eea
To illustrate the procedure with a specific example, we separate out the time coordinate and split the coordinates into
\bea
x^\m\equiv (t,y^a)
\eea
where $\m=0,..,3$ and $a=1,2,3$. (The resulting 3D system will be Euclidean, thus non-dynamical. For study of dynamics, one should consider a different setup by separating out, say, $y^3$ coordinate from the rest. We will come back to this issue below.)
By parameterizing the 4D metric \cite{Arnowitt:1962hi}\cite{Poisson}
\bea
\gh_{\m\n}=\left(
\begin{array}{cc}
-n^2+\g^{ab}N_{a} N_{ b} & N_{ a} \\
&\\
N_{ b} & \g_{ab}
\end{array}
\right)
\eea
the 3+1 splitting yields (the boundary terms will not be kept tract of)
\bea
S=\int d^4 x\;n\sqrt{-\g} \left(R^{(3)}-K^2+K_{ab}K^{ab}\right)
\la{1p3act}
\eea
with
\be
K_{ab}=\fr1{2n}\left(\mathscr{L}_{\pa_t} \g_{ab}-{\nabla}_a N_{b}
         -{\nabla}_b N_{ a} \right),\qquad K=\g^{ab}K_{ab}.
\la{K4defqq}
\ee
where $\mathscr{L}_{\pa_t}$ denotes the Lie derivative along time coordinate $t$ and $\N_a$ is the 3D covariant derivative (namely, its connection is constructed out of $\g_{ab}$); $n$ and $N_{ a}$ denote the lapse function and shift vector respectively. Since the time derivative does not act on $N_a$ or $n$ in their field equations, which read
\bea
{\N}_a (K^{ab}-\g^{ab} K)=0  \la{Ncon}
\eea
\bea
R^{(3)}+K^2-K_{ab}K^{ab}=0  \la{ncon}
\eea
these fields are non-dynamical: they do not have any time-derivative acting on them, and thus their bulk values can be taken as the corresponding values on the hypersurface of the fixed time once $n$ and $N_{ a}$ are specified on the hypersurface of a given time.
After the gauge-fixings that we turn to now, the equations above should be taken as the constraints that determine the physical states. (This is in the same spirit as the quantization of string (see, e.g., ch2 of \cite{GSW1}). The procedure is also consistent with the observation made in \cite{Kiriushcheva:2005yu}.) As we will see below, they can be combined and solved.

The action 
has the gauge symmetry of measure-zero (compared with the 4D gauge symmetry) left after imposing the 4D de Donder gauge. After we discuss gauge-fixing of the 4D diffeomorphism, we will come back to
this symmetry to fix $N_a=0$ on the hypersurface. (As far as we can see, this view is consistent with \cite{Isham:1984sb}\cite{Lee:1990nz}. Again the use of the ADM variables is crucial in isolating the unphysical fields: as shown in \cite{Kiriushcheva:2008fn} the first class constraints will generate the bulk diffeomorphism in the Hamiltonian approach with the full 4D metric chosen as the canonical variables.) Then by invoking the non-dynamical nature of $N_a$, the bulk value will be taken as its hypersurface value, namely, $N_a=0$ for the entire bulk. Similarly the lapse function can be fixed
to $n=1$ for the entire bulk.\footnote{{This choice will be good for quantization around a flat background. If one has a black hole background in mind with the radial coordinate separated out, one will have to introduce a radial coordinate transformation such that $g_{rr}$ does not depend on $r$. As we will further comment later, this is where one may encounter difficulties when trying to apply the procedure to an explicitly time-dependent background because it is not clear whether such a coordinate transformation will always be available without subtleties. This also reveals the background dependence of our procedure: although the methodology is background independent, the detailed steps will depend on the background under consideration. This aspect also makes a connection with precisely what physics the reduced theory describes, a point that we will discuss in section 3 and the conclusion.}}

Let us fix the 4D diffeomorphism by imposing the de Donder gauge. The full form of de Donder gauge is given by \cite{Smarr:1978dia}\footnote{As a matter of fact, one only needs the $\m=0$ part of this for the reduction:
 \bea
 \gh^{\r\s}\G_{\r\s}^0=0  \la{fulldd0}
 \eea
 The rest will serve as the 3D de Donder gauge in the reduced theory.
Improved versions of the reduction can be found in \cite{Park:2015qxa} and \cite{Park:2015ota}.
The account in \cite{Park:2015ota} is especially simple.  
 }
 \bea
 \gh^{\r\s}\G_{\r\s}^\m=0  \la{fulldd}
 \eea
{In the conventional perturbative analysis, one proceeds and adds the corresponding ghost term in the action. However, the measure-zero gauge fixing will affect, as we will show now, the bulk fixing simply because the bulk fixing \rf{fulldd} contains the non-dynamical fields that get fixed by the 3D gauge-fixing. At the end, only the 3D ghosts will be required after the system is reduced.}
In terms of the ADM variables, \rf{fulldd} translates to
 \bea
\gh^{\r\s} \G_{\r\s}^0=0 &:& (\pa_t-N^a \pa_a)n  = n^2 K \nn\\
\gh^{\r\s} \G_{\r\s}^a=0 &:& (\pa_t-N^b \pa_b)N^a =-n^2\Big[\g^{ab}\pa_b \ln n-\g^{bc}\G_{bc}^a\Big]   \la{ddadmour}
 \eea
where $\G_{bc}^a$ denotes the 3D Christoffel symbol. As well-known, de Donder gauge leaves residual symmetry (see for example \cite{Zee}; a nonlinear level discussion can be found in \cite{Park:2014noa}) although it is not manifest in \rf{ddadmour}. 
It is thus possible, by using the residual diffeomorphism\footnote{The ADM form of the action has the manifest 3D gauge symmetry. The (linear form of) 4D de Donder gauge has the same symmetry in the form of the residual symmetry that is parameterized by $\e_a(y)$. The shift vector can be gauge-fixed by using this 3D symmetry.}, to set
\bea
N_{a}=0,   \la{Nazero}
\eea
initially on the hypersurface of the fixed time. As mentioned above, this equation can then be taken as valid in the entire bulk due to the non-dynamism of $N_a$.
Substituting $N_{ a}=0$ into \rf{Ncon}, which now serves as a constraint, one gets
\bea
{\N}^a \left[\fr{1}{n}\Big(\mathscr{L}_{\pa_t} \g_{ab}
 -\g_{ab}\g^{cd}\mathscr{L}_{\pa_t} \g_{cd}\Big)\right]=0  \la{mtmconstr}
\eea
When the covariant derivative acts on objects other than $n$, it yields zero because the covariant derivative and the Lie derivative commute in the present case \cite{Kobayashi}. This step requires advanced-level knowledge in differential geometry, Ch.6 of \cite{Kobayashi}; details can be found in { \cite{Park:2014qoa,Park:2015qxa}}.
  The constraint \rf{mtmconstr} can be solved and implies
 \bea
 \pa_a n=0 \la{constronn}
 \eea
 Since $n$ is non-dynamical, namely, $n=n(y^a)$, \rf{constronn} implies that $n$ should be a constant\footnote{{Here what is meant by a ``constant" is a $y^a$-independent expression. In the case of expanding the 4D theory around a Schwarzschild background with the radial coordinate separated out, this step implies that $n$ should be independent of $(t,\th,\f)$.} }: the $N_a$ constraint implies
$n=n(t)$ but since $n$ is non-dynamical, one can set
\bea
n=1  \la{none}
\eea
even in the bulk.
This fixing of $n$ should be supplemented by its field equation which should now serve as a constraint
\bea
R^{(3)}+K^2-K_{ab}K^{ab}=0  \la{nconstr}
\eea
where $K_{ab}$ takes
\be
K_{ab}=\fr1{2}\mathscr{L}_{\pa_t} \g_{ab}
\ee
once \rf{Nazero} and \rf{none} are substituted into \rf{K4defqq}.
The constraint \rf{nconstr} allows one to rewrite action
\bea
S= \int d^4x\; \sqrt{-\g} \;R^{(3)}
\la{1p3actred}
\eea
where the overall factor 2 has been absorbed. The form of the
action \rf{1p3actred} does not yet imply that the system becomes three-dimensional.
This is because one should still consider the constraint \rf{nconstr}. We will
now argue that \rf{1p3actred} with \rf{nconstr} implies projection of the 4D dynamics onto 3D. But before we proceed, a cautionary remark is in order. As well known, 3D gravity does not have graviton. In this sense, our reduced system is not the genuine 3D gravity: it has two propagating degrees of freedom, the physical degrees of freedom of the original 4D gravity projected onto the hypersurface. (We will have more on this below.)

 Let us split the constraint \rf{nconstr} into two parts: the $R^\3$ term and the $K$ terms. They vanish separately: to see that $R^\3$ vanishes we just need to recall that it is an on-shell constraint. In other words, we define the physical states to be annihilated by \rf{nconstr}, in which the fields are now viewed as the corresponding
operators acting on the Fock space. What one can do is to consider the full field equation for the 3D metric (i.e., the one that follows from \rf{1p3actred}) and first obtain the mode expansion at the linear level.
Then the linear level expression can be substituted into the full field equation, from which one can obtain the full expansion by iteration. (For the definition of the physical states in the actual perturbative computations, one may expand the metric to the linear order and neglect the higher interactions. This is in the same spirit with the comment in one of the footnotes in section 15.7 of \cite{Weinberg2}.) The full expansion of
$\g_{ab}$ should yield zero once substituted into the $R^\3$ part of the constraint  \rf{nconstr}. With this, only the $K$ terms remain. As can be seen from the first equation in \rf{ddadmour} $K$ vanishes
\bea
K=0
\eea
leaving only
\bea
K_{ab}^2 =0  \la{Kabconstr}
\eea
{ (The same conclusion has been reached in a complementary analysis in \cite{Park:2015qxa} and \cite{Park:2015ota}.)} Since this is a positive definite metric, this implies
\bea
\mathscr{L}_{\pa_t} \g_{ab}=0
\eea
In other words, the physical states of the original 4D system are fully reduced to 3D. The projected system has two physical degrees of freedom: the metric of the hypersurface has six components. The first equation of \rf{ddadmour} imposes a constraint and the second equation of \rf{ddadmour} imposes the 3D de Donder gauge, effectively removing three degrees of freedom. (We will have more on this in the next section.)

In the remainder of this section, we discuss the case of separating out one of the spatial directions, say, $y^3$:
\bea
x^\m\equiv (z^m,y^3)\;\;\mbox{where}\;\; m=0,1,2
\eea
The procedure goes almost identically except several sign changes. For example, the split form of the action now takes
\bea
S=\int d^4 x\;n\sqrt{-\g} \left(R^{(3)}+K^2-K_{mn}K^{mn}\right)
\la{1p3actz}
\eea
with
\be
K_{mn}=\fr1{2n}\left(\mathscr{L}_{\pa_{y^3}} h_{mn}-{\nabla}_m N_{n}
         -{\nabla}_n N_{ m} \right),\qquad K=h^{mn}K_{mn}.
\la{K4defqq}
\ee
The only non-trivial difference compared with the $t$-splitting is the condition that corresponds to \rf{Kabconstr}:
\bea
K_{mn}^2 =0  \la{Kabconstry3}
\eea
Unlike in \rf{Kabconstr}, the contractions of indices in this equation are done with the 3D metric with $(-++)$ signature. In general, a Wick rotation must be considered in field theories with 
non-positive definite metrics since otherwise the path integral is not well-defined. In the gravitational case, the procedure has a subtlety addressed in \cite{Gibbons:1978ac}\cite{Schleich:1987fm}\cite{Mazur:1989by}. As far as we can see, it is not a subtlety associated with the Wick rotation but with the presence of unphysical component of the metric.
We will take up this issue in the next section; once the Wick rotation is carried out, \rf{Kabconstry3} leads to
\bea
K_{mn}=0,
\eea
namely, on-shell reduction to a dynamical 3D system.

\section{Quantization and renormalization \la{quan}}

The goal of the previous section was to remove unphysical degrees of freedom.
In this section we tackle the very issue of renormalizability of 4D Einstein gravity. Afterwards, we note that the holographic reduction admits a nice mathematical perspective for the cases of globally hyperbolic spacetimes.

For quantization, the $t$-separation case and $y^3$-separation case do not have an essential difference. Both cases are subject to the subtlety observed in \cite{Gibbons:1978ac,Schleich:1987fm,Mazur:1989by}, namely, the divergence associate with the trace part of the metric.

\subsection{4D quantization through hypersurface}

The upshot of the analysis in the previous section is that,
upon fixing $n=1, N_a=0$ by using lower dimensional diffeomorphism after the bulk fixing, the dynamics of the original 4D system is projected onto ``3D'':
\bea
S= \int d^4x\; \sqrt{-\g} \;R^{(3)}   \la{eff3Dact}
\la{1p3actredq}
\eea
The quotation mark around 3D above is put because of the presence of
the 4D integration and 4D coordinate dependence of the off-shell metric $\g_{ab}$. 
We now show that the integral over the 4D spacetime is effectively reduced to 3D. For the discussion in this section, we consider separating out the spatial coordinate $y^3$.
This action should be supplemented by the nonlinear
form of the 3D de Donder gauge
\bea
\g^{mn}\G_{mn}^p=0   \la{3dddz}
\eea
{With the genuine 3D gravity, one can use the residual symmetry after imposing \rf{3dddz} to gauge away three non-dynamical components, thereby arriving at well-known absence of propagating degrees of freedom. 
In the current case, however, it is impossible to do the same simply because the metric in \rf{eff3Dact} still has the 4th coordinate dependence off-shell.
The residual symmetry after \rf{3dddz} should be the symmetry within the hypersurface. This can be seen by considering the residual symmetry condition stated in \cite{Zee}. Namely, it is a partial differential equation of the gauge parameter $\e_{{m}}$ on the hypersurface and therefore cannot be used to gauge away additional components. Alternatively, one can consider the entire procedure of the gauge-fixings in the path-integral setup. 
The result \rf{eff3Dact} with \rf{3dddz} have been obtained as a result of this complete gauge-fixing by implicitly following the steps given in section 15.4 of \cite{Weinberg2}, and therefore one should not further gauge-fix \rf{eff3Dact} other than \rf{3dddz}.

{Remarks on precisely what physics the reduced system \rf{eff3Dact} with \rf{3dddz} is supposed to describe are in order. { The focus of the present work is perturbative quantization around a flat background, exactly the goal that we set out for. Although the procedure should be applicable to a Schwarzschild background, it is not clear whether the procedure can be used to study more complex backgrounds.} Related to this, there is an issue of non-perturbative corrections. We will not pursue these issues here.

\begin{figure}[tbp]
\centering 
\includegraphics[width=1.0\textwidth,trim=0 530 0 80,clip]{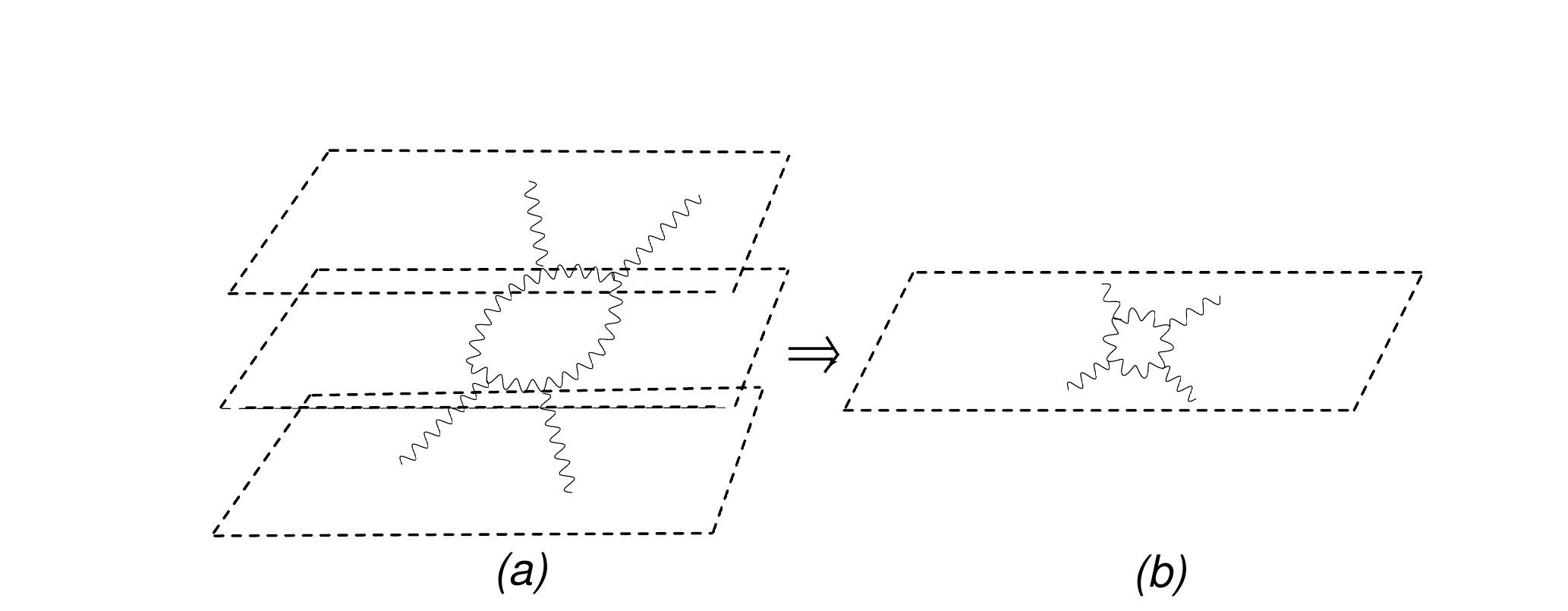}
\caption{\label{fig:i} (a) 4D scattering \;\;   (b) projection onto 3D hypersurface}
\end{figure}

{ For perturbation, one can use the usual linear form of the 3D de Donder gauge \rf{3dddz}. The 3D gauge-fixing and ghost terms will have to be introduced, and with that one can carry out the BRST quantization. Here we focus on the graviton part of the computation. Since the procedure maintains only the 3D covariance, one should worry about compatibility with the (yet unavailable) fully 4D covariant approach. We will further remark on this in the conclusion.}

The perturbation series with \rf{eff3Dact} can be set up as follows. { (The following analysis has been substantially expanded in \cite{Park:2014noa}.)} For the perturbative analysis, the usual linearized version of \rf{3dddz} can be imposed. The propagator is then given by
\bea
\!\!\!\!\!\!<q_{mn}(x_1)q_{pq}(x_2)>=\fr12 (\eta_{mp}\eta_{nq}+\eta_{mq}\eta_{np}-\fr23 \eta_{mn}\eta_{pq})\;\int \fr{d^4{k}}{(2\pi)^4} 
\fr{e^{i{k}\cdot ({x}_1-{ x}_2)}}{i\vec{k}^2}\nn\\
\eea
where $\vec{k}$ is the 3-vector part of the 4-momentum $k_\m$ and we have defined 
{
\bea
w\equiv y^3
\eea
}
The field $q_{rs}$ is the traceless part of the fluctuation around a flat metric (use of the traceless metric $q_{mn}$ is not essential for the reduction)
\bea
\g_{mn}\equiv \eta_{mn}+l_{mn},
\quad q_{mn}\equiv l_{mn}-\fr13 \eta_{mn}l,\quad l\equiv \eta^{mn}l_{mn}
\eea
This form of the propagator implies that one can carry out the renormalization program just as in the regular 4D case.
After performing various momentum integrals, thereby removing momentum delta functions, one will be left with integration over the loop momenta in which the propagators take the three-dimensional forms. The 3D integration can be carried out by the standard techniques. The integration over the reduced direction can be carried out by taking, e.g., momentum cut-off regularization. In general, momentum cut-off regularization does not respect gauge symmetry. However, adopting it for the 4th direction does not pose a problem here: dimensional regularization can be used for the 3D integrals. The divergence factor
arising from the $k^0$ integration can be absorbed by rescaling the fields (i.e., wave-function renormalization). In more detail, let us introduce a rescaling of $w$ by a dimensional parameter $L$:
\bea
u=\fr{w}{L}
\eea
such that $u$ is dimensionless.
With this rescaling, one can use a dimensionless momentum cut-off for the $u$ direction.}
The rescaling will lead to renormalization (,i.e., rescaling by $L$) of Newton's constant that has been suppressed, and one achieves at effective reduction to 3D:
\bea
S= \int d^3z\; \sqrt{-\g} \;R^{(3)}
\la{1p3actredq}
\eea 
Since the 4D system \rf{1p3act} with which we have started has reduced to \rf{1p3actredq}
with the constraint \rf{3dddz}, the former would be renormalizable if the latter is.

For renormalizability of \rf{1p3actredq}
with the constraint \rf{3dddz}, the result of \cite{Anselmi:2003xb} can mostly be borrowed with several cautions to which we will now turn.
 It was shown in \cite{Anselmi:2003xb} that 3D pure gravity is renormalizable even though it is power-counting non-renormalizable.
To use the result of \cite{Anselmi:2003xb} in the present case, one should first understand that the theory (or one of the theories) that was considered in \cite{Anselmi:2003xb} is the genuine 3D Einstein gravity that has no propagating degrees of freedom. In the case of the 3D pure gravity, the author relied on certain kinematic features of 3D gravity instead of explicit computation in order to deduce renormalizability.
However, the author considered propagating gravitons as can be seen in the gravity-matter coupling case where explicit computation was carried out. The propagating degrees of freedom must be the non-dynamical ones; this is in line with the common practice in the literature before and after \cite{Anselmi:2003xb} in which non-dynamical fields are allowed to propagate in loops. In the current setup, those non-dynamical fields are dynamical in the 4D standpoint, and are justified to be present in the loops.

The second cautionary step one should take is the fact that the result of \cite{Anselmi:2003xb} was obtained up to the observation in \cite{Mazur:1989by}, which addressed the issue raised in \cite{Gibbons:1978ac} and \cite{Schleich:1987fm}.
The constraint shows that the trace part of the original 4D metric is non-dynamical.
Upon the fixing of $n=1, N_a=0$, the constraint implies that the trace part of the
3D metric is non-dynamical. It was shown in \cite{Mazur:1989by} that the 4D trace part is non-dynamical in the 4D sense. Since our reduced system should represent the 4D system, the trace part of the metric (3D one after the fixing) should be excluded from the dynamics.

\subsection{ Perspective from mathematics of foliation }

The analysis of the previous section does not need any assumption on the causality of the starting 4D manifold. { It is possible to have an complementary understanding of the finding of the previous section through abstract foliation theory. We will assume for simplicity, although it is not necessary, that the background is globally hyperbolic.} This section is a brief summary of \cite{Park:2014qoa} in which more details can be found. (See \cite{Park:2015qxa} for further development of this line of the idea.)

A globally hyperbolic spacetime -which covers most of the cosmologically interesting spacetimes - admits a codimension-1 foliation through a family of hypersurfaces. 
The condition \rf{constronn} obtained by the shift vector constraint can be written precisely as the condition for the foliation to be Riemannian:
\bea
\mathscr{L}_{\pa_a} n=0  \la{Riemannconp}
\eea
In mathematics, it is know that a codimension-3 Riemannian foliation admits a ``dual" totally geodesic foliation. The duality involved is a mathematical one and operates between two different foliations: the Riemannian and totally geodesic.

Now the starting manifold can be viewed as a principle bundle of 1D abelian fibration over the 3D base through the totally geodesic foliation. Then one may associate the action of 1D group fibration with the gauge symmetry; modding out the gauge symmetry will correspond to casting the 4D manifold into its 3D base. { The presence of the totally geodesic foliation has an enlightening implication for the apparent gauge-choice sensitivity as we will discuss in the conclusion.}

\section{Conclusion}

A metric in four dimensions has ten components, four of which are non-dynamical, and another four components are associated with 4D gauge symmetry. In the conventional covariant renormalization program, only the modes associated with 4D gauge symmetry are removed by a gauge
fixing such as the de Donder gauge; the 4D Einstein gravity is power-counting unrenormalizable and indeed turned out unrenormalizable.
Even though it is only the measure-zero diffeomorphism that remains unfixed in the conventional covariant approach, the non-dynamical fields circulate the loop acting as {\em bulk} fields thereby  apparently ruining the renormalizability.

In this work, we have contemplated the possibility that additional removal of the non-dynamic fields may lead to renormalizability of {the ADM formulation of} 4D Einstein gravity. We have shown that there exists a way to remove all (or most) of the unphysical degrees of freedom and at the same time to set up a convenient stage for examining renormalizability. (As stated in one of the footnotes in the introduction, one may keep the 4 of the 8 unphysical fields at the off-shell level and carry out renormalization \cite{Park:2015ota}.) After removal of the unphysical fields, the 4D dynamics gets to admit the effective description in terms of the 3D language: 4D renormalizability is achieved based on the 3D renormalizability.

{Compared with AdS/CFT, it is notable that the boundary theory in the present case is not a gauge theory but takes a form of a lower dimensional gravity theory. We do not believe that this has anything to do the hypersurface foliation approach because the approach (combined with manual dimensional reduction) has led to a boundary {\em gauge} theory in the IIB supergravity setup \cite{Sato:2002kv}\cite{Hatefi:2012bp}.}

There is an interesting implication of the present work for the previous works of \cite{Park:2013iqa,Park:2013bma,Park:2014mba} wherein explicit dimensional reduction
was carried out in order to avoid the quantization issues. The analyses there were for certain sub-sectors (i.e., the sectors associated with the hypersurfaces selected) of the whole moduli space. The result of the present work implies that the analyses in those works are much more complete than originally believed.

\vspace{.1in}
We end with several future directions:
\vspace{.1in}

{ One of the subtle points is that the whole procedure of reduction seems to heavily depend 
on the elaborate gauge-fixings. (It would thus be worthwhile to explore whether there exists a bulk gauge other than the de Donder gauge that would lead to the same conclusion. Perhaps an easier task that will nevertheless shed light on this issue is to study how the form of the de Donder gauge \rf{fulldd} should be modified in the case of a Schwarzschild background.) Part of the reason for the gauge-sensitivity should be that the ratio of the physical versus unphysical fields is much higher than in YM theory.
It seems, therefore, natural to a certain extent that certain things depend on the gauge-choice more sensitively than in YM theory. The mathematical approach in { \cite{Park:2014qoa,Park:2015qxa}} - which takes advantage of the presence of the dual totally geodesic foliation - strongly suggests that the reduction should be a rather general phenomenon (associated with relatively simple backgrounds). This is because in that approach the only thing one needs is the very natural gauge-fixing, $\hat{g}_{0a}=0$ (which of course is $N_a=0$). Judging from this and other pieces of evidence, the gauge sensitivity of the present approach should not be such a serious problem.}

The present work was carried out for fluctuations around a flat spacetime. For example, it would definitely be worthwhile to carry out a similar analysis for a Schwarzschild black hole background. Although the methodology would be basically the same, the detailed steps would have to be modified: the present approach has moderate background dependence. To this end, it would be better to consider separating out the radial coordinate than the time coordinate. 
We believe that this is a task that can be carried out with only relatively minor changes in the present analysis. With this achieved, one would be in a good position to tackle the Firewall \cite{Almheiri:2012rt} (see \cite{Braunstein:2009my} for an earlier related work) and related issues, which was the main motivation for the present work.

{
At some point, one must face the difficult issue of non-perturbative contributions to the path-integral. There are two kinds of non-perturbative contributions. The first kind is the contribution coming from summing over all orders of the coupling constant in the perturbation theory. A 3D version of the techniques of the asymptotically safe gravity \cite{Weinberg3,Reuter:1996cp,Niedermaier:2006ns,Litim:2008tt,Percacci:2011fr} may be useful along this line. There will be several important issues such as the Gribov problem (see, e.g., \cite{Eichhorn:2013ug} and references therein for a recent account in the gravity context) that would have to be addressed eventually. 
The other kind is the contribution analogous to the instanton contribution in YM theory.
It is not clear whether the present procedure is adequate for that, given the limitation mentioned in the main body. (Unlike the instantons in YM theory, much less is known about the gravitational analogues anyway.)

 One of the utmost important directions concerns the 4D covariance since the present approach maintains only the 3D covariance. There exists several different ways of ensuring equivalence between a non-covariant approach and a covariant approach. (See \cite{Park:2015ota} for recent progress.) Let us recall by a few examples how the equivalence is accomplished in quantum field theory and string theory. One may attempt to construct 4D covariant operator formulation without using any constraint directly in the action. In other words, one would attempt 4D-covariant removal of all or most of the unphysical degrees of freedom. Perhaps it could be done along the line of the gauge invariant quantization \cite{Weinberg2}. {Some of the observations of \cite{Kiriushcheva:2008sf} may be useful.} While these programs could succeed, an easier approach will presumably be along the line of the illuminating analysis in section 9.6 of \cite{Weinberg1}. 
It was shown that the Coulomb gauge canonical quantization of electrodynamics can be related to the Lorentz gauge path integral. When both the non-covariant and covariant formulations were independently available, one could establish the equivalence by simply comparing the amplitudes of the physical states (and this is the way the equivalence between the lightcone quantization and covariant quantization was achieved in string theory). Computations of various Feynman diagrams must be preceded to this end.
Both of these approaches were discussed in \cite{Park:2014noa} at least to some extent.}

The observation on the renormalizability in this work is, to some extent, up to the issue of reduced space quantization vs. Dirac quantization. In general, those two approaches do not lead to the same physics \cite{Ashtekar:1982wv}\cite{Schleich:1990gd} (see therein for a more complete list of refs). This was demonstrated by taking an example in which the first constraint was not associated with the gauge symmetry. It was shown that, in general, both quantization methods lead to different but consistent theories. We are not aware of any work in which a similar conclusion was drawn for a theory with a gauge-symmetry inducing first class constraint. We believe that those two approaches are more likely than not to lead to the same results at the end as was the case in string quantization: light-cone, old covariant and modern BRST quantizations all led to the same results.

Finally, we have recently become aware of the works of York \cite{York:1972sj}, Moncrief \cite{Moncrief:1989dx} and Fischer-Moncrief \cite{Fischer:1996qg} wherein  certain Hamiltonian reductions were carried out on a certain class of 4D manifolds. It was our pleasant surprise to discover that the reduced Hamiltonians were given by the volumes of the hypersurfaces. Given that the action for a particle is given by its length and that of a string by its surface, the appearance of a volume for the hypersurface seems natural except that the volume appears as the Hamiltonian instead of the Lagrangian. Presumably this should be due the fact that the reduced direction is the time-direction so that the hypersurface is Euclidean.
In our case, it was the 3D Einstein action that has emerged. Perhaps, the 3D hypersurfaces admit dual descriptions, one through the volume and the other through the Einstein action.
It would be interesting to understand the potential relation between the works of \cite{York:1972sj}\cite{Moncrief:1989dx}\cite{Fischer:1996qg} and the present one.      
\\

We will report on progress made with some of these issues in the near future.\footnote{{Some of the issues have been recently addressed in \cite{Park:2014noa,Park:2015qxa,Park:2015ota}.}}

\vspace{.5in}
\ni {\bf Acknowledgments}

\ni I acknowledge the hospitality of S.-J. Sin and B.-H. Lee at Hanyang University and CQUeST during my stay. 
I thank the members of Hanyang University and KIAS for their interests and criticisms. This work benefited from the discussions I had with them. I thank V. Moncrief, G. 't Hooft, and R. Woodard for their communications.
I also thank H. Lee, E. Poisson, 
S. Weinberg for reading the main idea of the work and making encouraging comments.

\newpage

\appendix

\renewcommand{\theequation}{A.\arabic{equation}}
 \setcounter{equation}{0}
\section{Gauge-fixing dependence of renormalization}

{Let us demonstrate the complications caused by the presence of the unphysical modes in carrying out a renormalization program.  
We show that the renormalization procedure substantially depends on whether or not the undynamical metric component, the trace, is fixed.\footnote{Of course, keeping the trace part makes the path-integral ill-defined as observed in \cite{Gibbons:1978ac}, \cite{Schleich:1987fm} and \cite{Mazur:1989by}; for the sake of the discussion in this section, we set this observation aside and formally proceed for now. (This problem has been addressed recently in \cite{Park:2015ota}.)}  
As we will see, whether or not one gauge away the trace part makes the procedures drastically different. What we observed in the main body was that the difference can be so drastic as to render the 4D pure Einstein action renormalizable.

The gauge-fixing dependence of a renormalization procedure can easily be  illustrated by a coupled system of a metric and scalar.\footnote{The reduction to 3D that we observed in the main body was for the pure Einstein gravity; we do not claim the same for the coupled system that we consider in this section.}
Consider 4D Einstein action coupled with a scalar
\bea
S=\int d^4 x \sqrt{-\gh}\;(\Rh-\fr12 \gh^{\m\n}\pa_\m\f \pa_\n \f)
  \la{unsplitq}
\eea
and the weak-field expansion around Minkowski metric
\bea
\gh_{\m\n}=\eta_{\m\n}+g_{\m\n}
\eea
Let us impose de Donder gauge by adding the following gauge-fixing term to the action:
 \bea
 {\cal L}_{g.f.}&=&
     -{\fr12}(\pa_\r g^{\m\r}-\fr12 \pa^\m g)^2
 \label{ggf}
 \eea
where $g\equiv \eta^{\r\s} g_{\r\s}$.
Multiplying an overall factor 2, the action is given by
\bea
{\cal L}&=&  \Big(
 - \frac{1}{2}\del_\a g_{\m\n} \del^\a
g^{\m\n} { + \frac{1}{4} \del_\a g^\m_\m \del^\a g^\n_\n}
 \Big)-\pa^\n \Cb^\m \pa_\n C_\m \nn\\
&&-\fr12 g^{\a\b}\pa_\a g^{\g\d}\pa_\b g_{\g\d}+2g^{\a\b}\pa_\a g^{\g\d}\pa_\d g_{\b\g}
 { -g^{\a\b}\pa_\a g^{\e}_\e \pa_\d g^{\d}_\b-\fr12 g^{\a}_\a\pa_\d g^{\b\g}\pa_\g g_{\b\d}}  \nn\\
&&{  +\fr14 g^{\a}_\a\pa^\d g^{\b\g}\pa_\d g_{\b\g} -g^{\a\b}\pa_\d g^{\e}_\e\pa_\b g^{\d}_\a +\fr12 g^{\a\b}\pa_\a g^{\e}_\e\pa_\b g^{\r}_\r -g^{\a\b}\pa_\g g_{\a\b}\pa_\d g^{\g\d} }  \nn\\
&&{ +\fr12  g^{\e}_\e \pa_\g g^{\r}_\r \pa_\d g^{\g\d}+g^{\a\b}\pa^\g g_{\a\b}\pa_\g g^\e_{\e} +\fr14 g^\e_\e \pa_\g g^\r_\r \pa^\g g^\s_\s} -g^{\a\b} \pa_\d g^\g_\a \pa^\d g_{\b\g}  
+g^{\a\b}\pa_\d g^\g_{\a}\pa_\g g^\d_\b  \nn\\
&&
  - \pa^\n\Cb^\m \pa_\m C^\r g_{\n\r}
  -\pa^\n \Cb^\m \pa_\n C^\r g_{\m\r}
  -\pa^\n \Cb^\m C^\r\pa_\r g_{\m\n} \nn\\
  &&\;\;+\pa_\m \Cb^\m \pa^\n C^\r g_{\n\r}
  { +\fr12 \pa_\m \Cb^\m C^\r \pa_\r g^\n_\n }  \nn\\
&&-\fr12\Big[\eta^{\m\n}-g^{\m\n}+\fr12 g\eta^{\m\n}+g^{\m\r}g_{\r}^\n+\fr18 \eta^{\m\n}(g^2-2g_{\r\s}g^{\r\s})-\fr12 gg^{\m\n}\Big] \pa_\m\f
 \pa_\n \f\nn\\
 &&\quad+\cdots  
  \label{cubichzero}
\eea
where $C$ is the ghost field that corresponds to the gauge-fixing \rf{ggf}.
The trace part is not a dynamical degree of freedom \cite{Kuchar:1970mu}, and 
the traceless condition can explicitly be enforced by using the traceless propagator as follows.
Let us explicitly introduce the traceless metric
\bea
h_{\m\n}\equiv g_{\m\n}-\fr14 \eta_{\m\n}g
\la{htl}
\eea
and choose 
\bea
\cL_{kin}=-\fr12\pa_\a h_{\m\n}\pa^\a h^{\m\n}
\eea 
as the kinetic term. The propagator then is traceless:
\bea 
<h_{\m\n}(x)h_{\r\s}(y)>=\fr12(\eta_{\m\r}\eta_{\n\s}+\eta_{\m\s}\eta_{\n\r}-\fr12 \eta_{\m\n}\eta_{\m\s})  \int \fr{d^4k}{(2\pi)^4}\fr{e^{ik\cdot (x-y)}}{i k^2}
\eea
It follows from this
\bea
<h_{\m\n}(x)h(y)>=0=<h(x)h_{\r\s}(y)>;
\eea
any vertex term that contains a factor $h$ will lead to zero in correlator computation. All the quadratic terms containing a $h$-factor will not contribute to any correlator, and thereby be effectively dropped; the action becomes substantially simplified:
\bea
\cL &=&
 - \frac{1}{2}\del_\a h_{\m\n} \del^\a h^{\m\n}
 -\pa_\n \Cb_\m \pa^\n C^\m \nn\\
&&{ -\fr12\Big[}-\fr12 h^{\a\b}\pa_\a h^{\g\d}\pa_\b h_{\g\d}+2h^{\a\b}\pa_\a h^{\g\d}\pa_\d h_{\b\g}
 -h^{\a\b} \pa_\d h^\g_\a \pa^\d h_{\b\g}  \nn\\
&&\hspace{2in}+h^{\a\b}\pa_\d h^\g_{\a}\pa_\g h^\d_\b  { -h^{\a\b}\pa_\g h_{\a\b}\pa_\d h^{\g\d} } { \Big]}\nn\\
&&
  - \pa^\n\Cb^\m \pa_\m C^\r h_{\n\r}
  -\pa^\n \Cb^\m \pa_\n C^\r h_{\m\r}
  -\pa^\n \Cb_\m C^\r\pa_\r h_{\m\n} +\pa_\m \Cb^\m \pa^\n C^\r h_{\n\r}
  \nn\\
  &&-\fr12\Big[\eta^{\m\n}-h^{\m\n}+h^{\m\r}h_{\r}^\n-\fr14 \eta^{\m\n}h_{\r\s}h^{\r\s}\Big] \pa_\m\f
 \pa_\n \f+\cdots     \la{haction}
\eea
One can check that $h=0$ is a legitimate gauge. At one-loop (see Fig. 1 (a)), $<\f(x_1)\f(x_2)\f(x_3)\f(x_4)>$, which is a gauge-singlet, is same whether one uses \rf{cubichzero} or \rf{haction}. 

\bea
&&\begin{tikzpicture}[line width=1 pt, scale=1]
	\draw (-140:1)--(0,0);
	\draw (140:1)--(0,0);
   \draw[vector] (.52,0) circle (.47cm);	

	\node at (-140:1.2) {$\f$};
	\node at (140:1.2) {$\f$};
		
\begin{scope}[shift={(1,0)}]
	\draw (-40:1)--(0,0);
	\draw (40:1)--(0,0);
	\node at (-40:1.2) {$\f$};
	\node at (40:1.2) {$\f$};	
	\node at (70:-1.2) {(a)};	
\end{scope}
\begin{scope}[shift={(5.0,0)}]
			\draw (-140:1)--(0,0);
			\draw (140:1)--(0,0);
			\draw[vector] (0:1)--(0,0);
			\node at (-140:1.2) {$\f$};
			\node at (140:1.2) {$\f$};
			\node at (90:-1.2) {$(b)$};		
		
\end{scope}
\end{tikzpicture}
\hspace{.5in}
\begin{tikzpicture}[line width=1.0 pt, scale=.65]
     \begin{scope}[rotate=90]	
		\draw (-.5,2.3) -- (0,2);
		\draw[vector] (0,2) -- (2,2);
		\draw(2,2) -- (2.4,2.4);
		
		\draw (0,2) -- (1,.5); 
		\draw (1,.5) -- (2,2);
		
		\draw[vector] (1,.5) -- (1,-.5);
		\node at (2.5,2.7) {$\f$};
		\node at (-.6,2.6) {$\f$};
\node at (-1,.8) {$(c)$};
	\end{scope}
	\end{tikzpicture}\nn\\
&&\hspace{1.0in}\mbox{ Figure 2: tree and one-loop diagrams	}  \nn
\eea
The see the gauge-dependence, let us consider scalar-scalar-metric amplitudes drawn in Fig. (b),(c). One can easily see that the tree level correlators $<\f\f g_{\m\n}>$ with the action \rf{cubichzero} and $<\f\f h_{\m\n}>$ with the action \rf{haction} lead to different results.
The same is true for the one-loop correlators. The computations are simplified in the latter case because the number of relevant vertices are smaller. Although we have not explicitly carried out, the diagram in Fig.2 clearly shows the drastic simplification in the computation with $h_{\m\n}$ as can be seen by counting the number of relevant vertices: 13 in \rf{cubichzero} vs 5 in \rf{haction}.  
\bea
\begin{tikzpicture}[line width=1.0 pt, scale=.65]
     \begin{scope}[rotate=90]	
		\draw (-.5,2.3) -- (0,2);
		\draw (0,2) -- (2,2);
		\draw(2,2) -- (2.4,2.4);
		
		\draw[vector] (0,2) -- (1,.5); 
		\draw[vector] (1,.5) -- (2,2);
		
		\draw[vector] (1,.5) -- (1,-.5);
		\node at (2.5,2.7) {$\f$};
	
		\node at (-.6,2.6) {$\f$};
\node at (-2,.8) {Figure 3: diagram with cubic metric coupling};
	\end{scope}
	\end{tikzpicture}
\eea
Carrying out a normalization procedure for $g_{\m\n}$ means that we consider the amplitudes in which the unphysical component $g$ appears, e.g., as an external line. It seems highly likely that exclusion of $g$ 
will reduce the number of counter terms. We have confirmed this expectation in the main body in a very dramatic way.}

%

\end{document}